\documentclass[10pt,preprint]{sigplanconf}

\usepackage{epsfig}
\usepackage{boxedminipage}
\usepackage{latexsym}
\usepackage{balance}

\usepackage{amsmath}
\usepackage{amsthm}
\usepackage{amssymb}
\usepackage{hyperref}
\usepackage{graphicx}
\usepackage{subfigure}
\usepackage[english]{babel}
\usepackage[utf8x]{inputenc}
\usepackage{eso-pic}
\usepackage{alltt}
\usepackage{listings}
\usepackage{algorithm}
\usepackage{algorithmic}
\usepackage{framed}
\usepackage{float}
\usepackage{subfloat}
\usepackage{wrapfig}
\usepackage{multirow}
\definecolor{shadecolor}{gray}{0.9}

\usepackage{blindtext}

\usepackage{etoolbox}
\makeatletter
\patchcmd{\maketitle}{\@copyrightspace}{}{}{}
\makeatother

\def\fullpaper#1{}

\newcommand{\stt}{\small\tt}

\newcommand{\rms}{{\sc rms}}
\newcommand{\trms}{{\sc trms}}

\newtheorem{definition}{Definition}

\begin{document}

\conferenceinfo{WXYZ '13}{date, City.} 
\copyrightyear{2013} 
\copyrightdata{[to be supplied]} 

\newcommand{\emilio}{{$\bullet \bullet \bullet $}}
\newcommand{\camil}{{$\bullet \bullet \bullet $}}
\newcommand{\irene}{{$\bullet \bullet \bullet $}}
\newcommand{\romolo}{{$\bullet \bullet \bullet $}}

\title{Multithreaded Input-Sensitive Profiling}

\authorinfo{Emilio Coppa}
           {{\small Dept. of Computer Science\\Sapienza University of Rome}}
           {coppa@di.uniroma1.it}
\authorinfo{Camil Demetrescu}
           {{\small Dept. of Computer and System Sciences\\Sapienza University of Rome}}
           {demetres@dis.uniroma1.it}
\authorinfo{Irene Finocchi}
           {{\small Dept. of Computer Science\\Sapienza University of Rome}}
           {finocchi@di.uniroma1.it}
\authorinfo{Romolo Marotta}
           {{\small Dept. of Computer and System Sciences\\Sapienza University of Rome}}
           {romolo.marotta@gmail.com}

%\authorinfo{\camil\ \camil\ \hspace{3mm} \camil\ \camil\ \hspace{3mm} \camil\ \camil\ \hspace{3mm} \camil\ \camil}
%           {\camil\ \camil\ \camil\ \camil\ \\ \camil\ \camil\ \camil\ \camil}
%           {\camil\ \camil\ \camil}

\maketitle

\begin{abstract}
% !TEX root = paper.tex
Input-sensitive profiling is a recent performance analysis technique that makes it possible to estimate the empirical cost function of individual routines of a program, helping developers understand how performance scales to larger inputs and pinpoint asymptotic bottlenecks in the code. A current limitation of input-sensitive profilers is that they specifically target sequential computations, ignoring any communication between threads. In this paper we show how to overcome this limitation, extending the range of applicability of the original approach to multithreaded applications and to applications that operate on I/O streams. We develop new metrics for automatically estimating the size of the input given to each routine activation, addressing input produced by non-deterministic memory stores performed by other threads as well as by the OS kernel (e.g., in response to I/O or network operations). We provide real case studies, showing that our extension allows it to characterize the behavior of complex applications more precisely than previous approaches. An extensive experimental investigation on a variety of benchmark suites (including the SPEC OMP2012 and the PARSEC benchmarks) shows that our Valgrind-based input-sensitive profiler incurs an overhead comparable to other prominent heavyweight analysis tools, while collecting significantly more performance points from each profiling session and correctly characterizing both thread-induced and external input.

%Input-sensitive profiling is a recent performance analysis technique that allows it to estimate the empirical cost function of individual routines of a program, helping developers pinpoint asymptotic bottlenecks in the code. In this paper we design and implement new metrics and algorithms for input-sensitive profiling. The goal is to support multithreaded applications and applications that operate on I/O streams, significantly extending the range of applicability of the original approach. We develop new metrics for estimating the size of the input given to each routine activation, including non-deterministic memory stores performed by the OS kernel, e.g., in response to I/O operations, or by other threads. We provide several case studies showing that our extension allows it to characterize the behavior of complex applications more precisely than the previous approaches, studying how the performance of each thread scales as a function of the amount of data received as input from other threads and from the OS kernel. An extensive experimental investigation shows that our extension incurs a small overhead, while collecting significantly more performance points from each profiling session in different scenarios.

\end{abstract}

%\category{D.2.2}{Software Engineering} {Tools and Techniques}[programmer workbench]
%\category{D.2.5}{Software Engineering} {Testing and Debugging}[diagnostics, tracing]


% !TEX root = paper.tex
%--------------------------------------------------------------------------------
\section{Introduction}
\label{se:intro}

Performance profilers collect information on running applications and associate performance metrics to software locations such as routines, basic blocks, or calling contexts~\cite{GKM82,S04,ABL97}. They play a crucial role towards software comprehension and tuning, letting developers identify hot spots and guide optimizations to portions of code that are responsible of excessive resource consumption. 

Unfortunately, by reporting only the overall cost of portions of code, traditional profilers do not help programmers to predict how the performance of a program scales to larger inputs. To overcome this limitation, some recent works have addressed the problem of designing and implementing performance profilers that return, instead of a single number representing the cost of a portion of code, a cost function that relates the cost to the input size (see, e.g.,~\cite{CDF12,GAW07,ZH12}). This approach is inspired by traditional asymptotic analysis of algorithms, and makes it possible to analyze -- and sometimes predict -- the behavior of actual software implementations run on deployed systems and realistic workloads. Some of the proposed methods, such as~\cite{GAW07}, perform multiple runs with different and determinable input parameters, measure their cost, and fit the empirical observations to a model that predicts performance as a function of workload size. More recent approaches make a step further, tackling the problem of automatically measuring the size of the input given to generic routines~\cite{CDF12,ZH12}, collecting data from multiple or even single program runs. 

As observed in~\cite{CDF12} and~\cite{ZH12}, a current limitation of input-sensitive profilers is that they specifically target sequential computations, ignoring any communication between threads. Multithreaded applications based on concurrent programming are traditionally difficult to analyze, since threads can interleave in a nondeterministic fashion and affect the behavior of other threads. Nevertheless, they are widespread in modern multicore architectures, making the quest for dynamic analysis tools for concurrent computations extremely critical. 

\paragraph{Our contribution.} 
In this paper we show how to extend the input-sensitive profiling methodology to the full range of concurrent applications, hinging upon the approach described in~\cite{CDF12}. The ability to automatically infer the size of the input data on which each routine activation operates is a crucial issue in input-sensitive profiling, but current techniques may fail to properly characterize the input size in a multi-threaded environment. As shown in this paper, if the input size is not estimated correctly, the analysis of profiling data can lead to  uninformative cost plots or even to misleading results. As a first contribution we therefore propose a novel metric, called {\em threaded read memory size}, that overcomes this limitation, addressing input produced by non-deterministic memory stores performed by other threads and by the OS kernel (e.g., in response to I/O or network operations). We provide real case studies, based on the {\tt MySQL} database management system and on the {\tt vips} image processing tool, showing that our extension allows it to characterize the behavior of complex applications more precisely than previous approaches. 
We then demonstrate that the input size of a routine can be automatically and efficiently computed  in a multithreaded setting, and discuss the implementation of a Valgrind-based input-sensitive profiler for concurrent applications (the tool is available at {\stt http://code.google.com/p/aprof/}). An extensive experimental investigation on a variety of benchmark suites (including the SPEC OMP2012 and the PARSEC benchmarks) shows that our profiler incurs an overhead comparable to other prominent Valgrind tools, while collecting significantly more performance points from each profiling session and correctly characterizing both thread-induced and external input.

%(Background on input-sensitive profiling: non abbiamo pi\'u la parte sulle tuple)

\paragraph{Paper organization.} 
The remainder of this paper is organized as follows. 
In Section~\ref{se:rvms} we introduce the threaded read memory size metric, showing its usefulness through synthetic examples. Case studies drawn from real applications are discussed in Section~\ref{se:case-studies}. Section~\ref{se:algorithms} proposes an efficient algorithm for computing the threaded read memory size of a routine activation. Section~\ref{se:implementation} describes the most relevant implementation aspects and Section~\ref{se:experiments} presents the results of our experimental evaluation. Related work is discussed in Section~\ref{se:related} and concluding remarks are given in Section~\ref{se:conclusion}.

% !TEX root = paper.tex
%--------------------------------------------------------------------------------
\section{Multithreaded Input Size Estimation}
\label{se:rvms}

A crucial issue in input-sensitive profiling is the ability to automatically infer the size of the input data on which each routine activation operates. This can be done in a single-threaded scenario using the {\em read memory size} metric introduced in~\cite{CDF12}:

\begin{definition}[\cite{CDF12}]
\label{de:rms}
The {\em read memory size} (\rms) of the execution of a routine $r$ is the number of distinct memory cells first accessed by $r$, or by a descendant of $r$ in the call tree, with a read operation.
\end{definition}

\noindent The intuition behind this metric is the following. Consider the first time a memory location $\ell$ is accessed by a routine activation $r$: if this first access is a read operation, then $\ell$ contains an input value for $r$. Conversely, if $\ell$ is first written by $r$, then later read operations will not contribute to increase the \rms\ since the value stored in $\ell$ was produced by $r$ itself. 

%the value contained in $\ell$ is regarded as input for $r$  if and only if the first access is a read operation.  

\begin{figure}[t]
\centering
\begin{tabular}{cc}
~ & ~ \\
\hspace{-4mm}{\raisebox{9mm}{\small{(a)}}} & \includegraphics[clip=true, height=1.80cm]{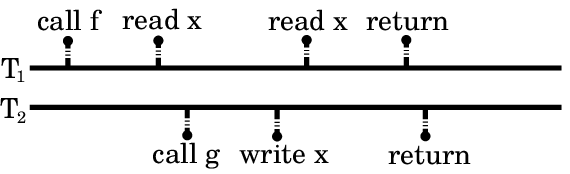} \\
~ & ~ \\
\hspace{-4mm}{\raisebox{8mm}{\small{(b)}}} & \hspace{-2mm}\includegraphics[clip=true, height=1.80cm]{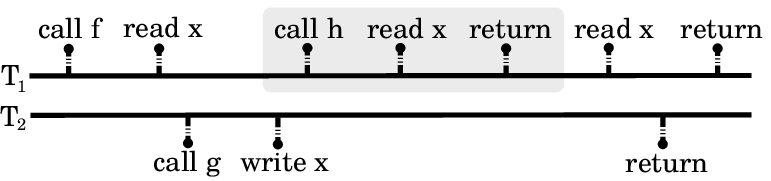} 
\end{tabular}
\caption{Threaded read memory size examples.}
\label{fig:trms-examples}
\end{figure}

The \rms, although very effective in single-threaded executions, may fail to properly characterize the input size of routine activations in a multi-threaded environment. Consider, as an example, the execution described in Figure~\ref{fig:trms-examples}a: routine $f$ in thread $T_1$ reads location $x$ twice, but only the first read operation is a first access. Hence, \rms$_f=1$. Notice however that routine $g$ in thread  $T_2$ overwrites the value stored in $x$ before the second read by $f$: this read operation gets a value that is not produced by routine $f$ itself and that should be therefore regarded as new input to $f$. The same drawbacks discussed in the above example arise when one or more memory locations are repeatedly loaded by a routine with values read from an external source, e.g., network or secondary storage. To overcome these issues, we propose a novel metric for estimating the input size, which we call {\em threaded read memory size}.

\begin{definition}
Let $r$ be a routine activation by thread $t$ and let $\ell$ be a memory location. An operation on $\ell$ is:
\begin{itemize}
\item  a {\em first-access}, if $\ell$ has never been accessed before by $r$ or by any of its descendants in the call tree;

\item an {\em induced first-access}, if  the  latest {\tt write}$(\ell)$ performed by any thread $t'\neq t$, if any, has not been followed by an access to $\ell$ by routine $r$ or by any of its descendants in the call tree.
\end{itemize}
\end{definition}

\begin{definition}
Let $r$ be a routine activation by thread $t$. The {\em threaded read memory size} \trms$_{r,t}$ of $r$ with respect to $t$ is the number of read operations performed by $r$ that are first-accesses or induced first-accesses.
\end{definition}

\noindent We notice that the \rms\ coincides with the number of read operations that are first-accesses and therefore 
\begin{equation}
\label{eq:trms-larger}
\mbox{\trms}_{r,t}\ge\mbox{\rms}_{r}
\end{equation}
for each routine activation $r$ and thread $t$. 

\paragraph{Example 1.} Consider again the example in Figure~\ref{fig:trms-examples}a: we have \trms$_{f,T_1}=2$. The first read operation on $x$ is indeed a first-access, while the second one is an induced first-access: between the latest write operation on $x$ performed by thread $T_2\neq T_1$ and the second {\tt read}$(x)$ by routine $f$ there are no other accesses to $x$ by $f$.

\paragraph{Example 2.} Consider Figure~\ref{fig:trms-examples}b. In this case \rms$_{h}=1$ and \rms$_{f}=1$: function $f$ performs three read operations on $x$ (one of which through its subroutine $h$), but only the first one is a first-access and contributes to its \rms. With respect to the \trms$_{f,T_1}$, the read operation by $h$ is an induced first-access for $f$ (similarly to the previous example), while the third read is not: between the latest write operation on $x$ performed by thread $T_2\neq T_1$ and the third {\tt read}$(x)$, $f$ has already accessed $x$ through its descendant $h$.  

We also have \trms$_{h,T_1}=1$. Notice that the read operation in $h$ could be regarded both as a first-access and as an induced first-access with respect to $h$: since we are interested in characterizing communication between threads via shared memory, we will classify accesses of this kind as induced first-accesses.

\begin{figure}[t]
\centering
%-----------------------------------------------
\begin{minipage}[t]{0.50\linewidth}
\flushleft
\textbf{procedure} {\tt producer()}
\vspace{-.5cm}
\begin{algorithmic}[1]
\STATE \textbf{while} $(1)$ \textbf{do}
\STATE ~~~~{\tt wait($empty$)}
\STATE ~~~~{\tt wait($mutex$)}
\STATE ~~~~{\tt produceData($x$)}
\STATE ~~~~{\tt signal($mutex$)}
\STATE ~~~~{\tt signal($full$)}
\end{algorithmic}
\end{minipage}
%-----------------------------------------------
\begin{minipage}[t]{0.49\linewidth}
\flushleft
\textbf{procedure} {\tt consumer()}
\vspace{-.5cm}
\begin{algorithmic}[1]
\STATE \textbf{while}  $(1)$ \textbf{do}
\STATE ~~~~{\tt wait($full$)}
\STATE ~~~~{\tt wait($mutex$)}
\STATE ~~~~{\tt consumeData($x$)}
\STATE ~~~~{\tt signal($mutex$)}
\STATE ~~~~{\tt signal($empty$)}
\end{algorithmic}
\end{minipage}
\vspace{.1cm}
\caption{Producer-consumer pattern: when $n$ values have been produced, \rms$_{\tt consumer}=1$ while \trms$_{\tt consumer}=n$.}
\label{fig:producer-consumer}
\end{figure} 

\paragraph{Example 3.} Producer-consumer is a classical pattern in concurrent applications. The standard implementation based on semaphores (see, e.g.~\cite{tanenbaum06}) is shown in Figure~\ref{fig:producer-consumer}, where producer and consumer run as different threads and routines {\tt produceData} and {\tt consumeData} write to and read from memory location $x$, respectively (the implementation can be easily extended to buffered read and write operations). For simplicity of exposition, we will not consider memory accesses due to semaphore operations. With this assumption, \rms$_{\tt consumer}=1$, since the consumer repeatedly reads the same memory location $x$. Conversely, the threaded read memory size gives a correct estimate of the consumer's input size: whenever {\tt producer} has generated $n$ values written to location $x$ at different times, we have \trms$_{\tt consumer}=n$. Indeed, all read operations on $x$ are induced first-accesses: thanks to the interleaving guaranteed by semaphores, each {\tt read}$(x)$ in {\tt consumeData} is always preceded by a {\tt write}$(x)$ in {\tt produceData}.

\begin{figure}[t]
\centering
%-----------------------------------------------
\begin{minipage}[t]{0.99\linewidth}
\flushleft
\textbf{procedure} {\tt externalRead()}
\vspace{-.5cm}
\begin{algorithmic}[1]
\STATE let $b$ a buffer of size $2$
\STATE \textbf{for} $i = 1$ to $n$ \textbf{do}
\STATE ~~~load $b$ with external data~~~~// does not imply read of $b$
\STATE ~~~{\tt consumeData($b[0]$)}~~~~~~~~~// read and process $b[0]$
\end{algorithmic}
\end{minipage}
\vspace{.1cm}
\caption{Buffered read from an external device: after $n$ iterations, \rms$_{\tt externalRead}=1$ and \trms$_{\tt externalRead}=n$.}
\label{fig:irene}
\end{figure}

\paragraph{Example 4.} The example in Figure~\ref{fig:irene} describes the case of buffered read operations. Procedure {\tt externalRead} loads $2n$ values from an external device (line 3): this is done by the operating system that fills in buffer $b$ with new data at each iteration. These load operations, however, should not be implicitly regarded as read operations performed by the running thread: as shown in line 4, only one of the two values loaded at each iteration is actually read and processed by procedure {\tt externalRead}. Hence, at the end of the execution \trms$_{\tt externalRead}=n$, due to the $n$ induced first-accesses at line 4. Notice that \rms$_{\tt externalRead}=1$ since data items are loaded across iterations on the same two memory locations $b[0]$ and $b[1]$ but only $b[0]$ is repeatedly read. We will further discuss the interaction between kernel system calls and threads in Section~\ref{ss:externalInput}.

% !TEX root = paper.tex
%--------------------------------------------------------------------------------
\section{Case Studies}
\label{se:case-studies}

In this section we discuss the utility of the \trms\ metric in real applications. We show a variety of cases where \trms\ correctly characterizes the input size where \rms\ either fails or does not collect sufficient profiling data. Our examples are based on the {\tt aprof-trms} tool described in Section~\ref{se:implementation} and use basic block (BB) counts as performance metric.

Input sensitive profiles can be naturally used to produce performance charts where some cost measure is plotted against the \trms\ or the \rms. For instance, for each distinct input size $n$ of a routine $r$, we can plot the maximum time spent by an activation of $r$ on input size $n$ (worst-case running time plot) or the number of times $r$ was activated on an input of size $n$ (workload plot). Similar charts could be produced for different cost measures (e.g., average running time), though we will not use them throughout this section.

Our discussion is based on two different applications: {\tt MySQL}, a relational database management system~\cite{MySQL}, and {\tt vips}, an image processing software package included in the PARSEC 2.1 benchmark suite~\cite{BKSL08}. {\tt MySQL} manages every new connection to the database by means of a separate thread, which contends for access to different shared data structures, and uses both  I/O and network intensively.  We also remark that {\tt vips} is a data-parallel application, which constructs multi-threaded image processing pipelines in order to apply fundamental image operations such as affine transformations and convolutions.

\begin{figure}[t]
\centering
\begin{tabular}{cc}
\hspace{-3mm}
\includegraphics[width=0.445\columnwidth]{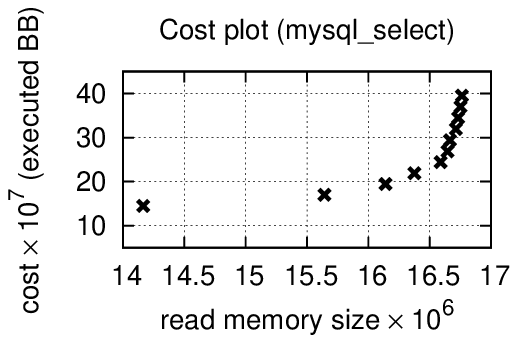} &
\includegraphics[width=0.445\columnwidth]{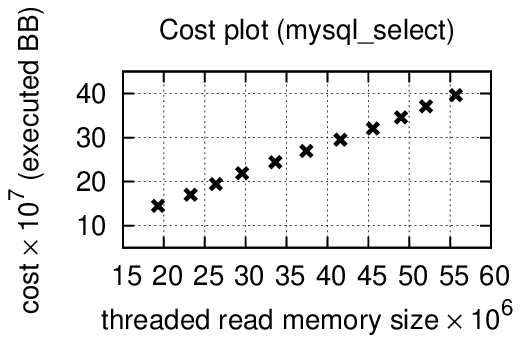}\hspace{-2mm} \vspace{-2mm}\\
\hspace{3mm}\small{(a)} & \hspace{4mm}\small{(b)} 
\end{tabular}
\caption{Function {\tt mysql\_select} of {\tt MySQL}: worst-case running time plots respectively obtained using \rms\ or \trms\ as an estimate for  the input size.}
\label{fig:mysql-select}
\end{figure}

\begin{figure}[t]
\centering
\begin{tabular}{cc}
\hspace{-3mm}
\includegraphics[width=0.445\columnwidth]{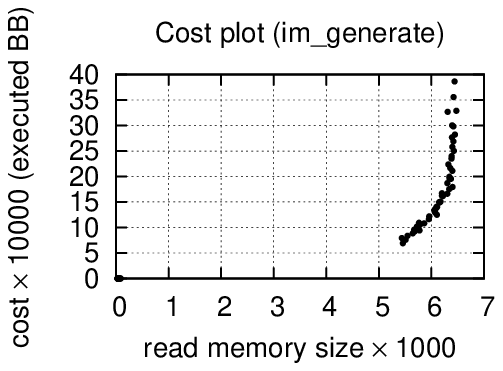} &
\includegraphics[width=0.445\columnwidth]{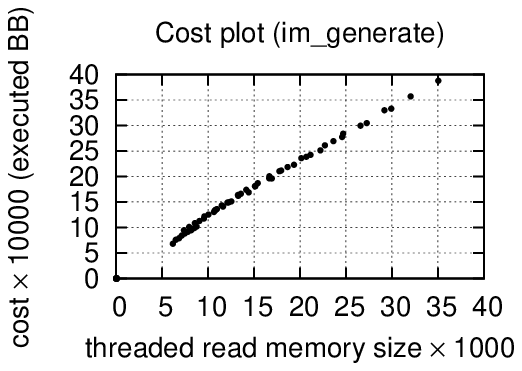}\hspace{-2mm} \vspace{0mm}\\
\hspace{3mm}\small{(a)} & \hspace{4mm}\small{(b)} 
\end{tabular}
\caption{Function {\tt im\_generate} of {\tt vips} ({\small PARSEC 2.1}): worst-case running time plots respectively obtained using \rms\ or \trms\ as an estimate for  the input size.}
\label{fig:vips-im-generate}
\end{figure}

%--------------------------------------------------------------------------------
\paragraph{Impact of input size estimation on asymptotic trends.}

If the input size is not estimated correctly, the analysis of profiling data can lead to misleading results. Consider for instance the following scenario: we have $n$ activations $r_1 ...\,r_n$ of a routine $r$, activation $r_i$ has cost $i$ and performs $i$ read operations, out of which $\lceil i/2\rceil$ are first-accesses and $\lfloor i/2\rfloor$ are induced first-accesses. Hence, $\mbox{\trms}_{r_i}=\lceil i/2\rceil+\lfloor i/2\rfloor=i$ while $\mbox{\rms}_{r_i}=\lceil i/2\rceil$. Notice that $\mbox{\trms}_{r_i}\ge\mbox{\rms}_{r_i}$, in accordance with Inequality~\ref{eq:trms-larger}. In the worst-case running time plot obtained using the \trms\ we have $n$ distinct points and the running time grows as the function $f(x)=x$. Conversely, in the worst-case running time plot obtained using the \rms\ we have only $n/2$ points: any two consecutive activations $r_i$ and $r_{i+1}$ (with $i$ odd) have the same \rms\ value $\lceil i/2\rceil$ and the worst-case cost is $i+1$ (i.e., the maximum between costs $i$ and $i+1$ of the two activations). Hence, 
the running time appears to grow as the function $f(x)=2x$. The problem would be even more critical if, e.g., $\mbox{\rms}_{r_i}=\lfloor\log i\rfloor$: in this case, in the worst-case  plot obtained using the \rms, the running time would appear to grow exponentially as $f(x)=2^x$.

\begin{figure}[t]
\centering
\begin{tabular}{cc}
\hspace{-3mm}
\includegraphics[width=0.445\columnwidth]{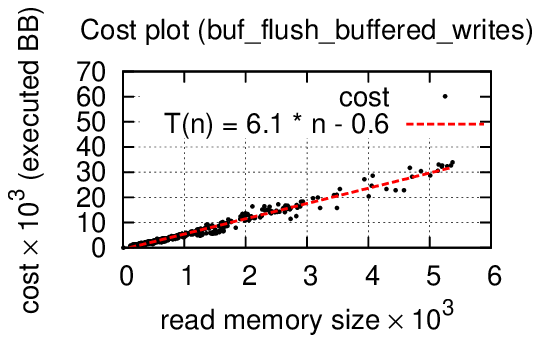} &
\includegraphics[width=0.445\columnwidth]{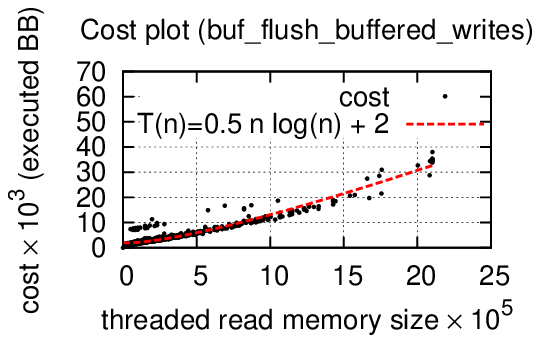}\hspace{-2mm}\vspace{-2mm} \\
\hspace{3mm}\small{(a)} & \hspace{4mm}\small{(b)} 
\end{tabular}
\caption{Function {\tt buf\_flush\_buffered\_writes} of {\tt MySQL}: worst-case running time plots with curve fitting.}
\label{fig:mysql-buf-flush}
\end{figure}

\begin{figure*}[ht]
\centering
\begin{tabular}{cccccc}
\hspace{-3mm}
{\raisebox{12mm}{\small{(a)}}} & \includegraphics[width=0.450\columnwidth]{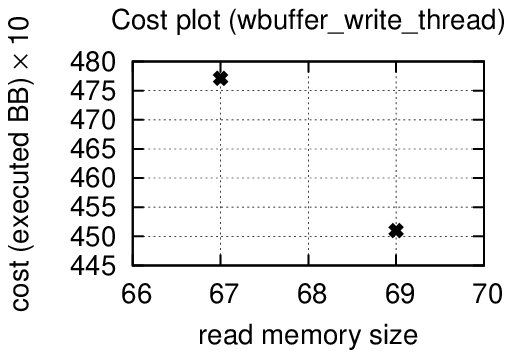}\hspace{+5mm} &
{\raisebox{12mm}{\small{(b)}}} & \includegraphics[width=0.450\columnwidth]{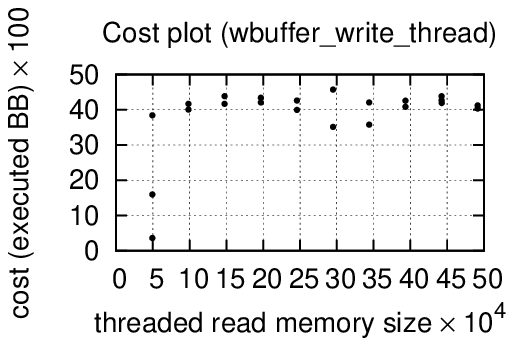}\hspace{+5mm} &
{\raisebox{12mm}{\small{(c)}}} & \includegraphics[width=0.450\columnwidth]{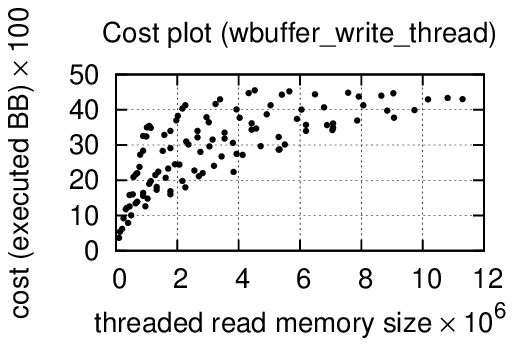}
%\hspace{4mm}\small{(a)} & \hspace{0mm}\small{(b)} & \hspace{0mm}\small{(c)} 
\end{tabular}
\caption{Function {\tt wbuffer\_write\_thread} of {\tt vips} ({\small PARSEC 2.1}): (a) \rms\ cost plot; (b) \trms\ cost plot with external input only; (c) \trms\ cost plot with both external and thread input.}
\label{fig:vips-wbuffer-write-thread}
\end{figure*} 

As shown by figures~\ref{fig:mysql-select},~\ref{fig:vips-im-generate},  and~\ref{fig:mysql-buf-flush}, similar phenomena can arise in practice in I/O bounded or multithreaded applications. The running time of routine {\tt mysql\_select} in Figure~\ref{fig:mysql-select} appears to grow (at least) quadratically when we measure the input size by means of the \rms, and linearly using the \trms. In this experiment, the query operation simply selects all tuples in the table and is applied to tables of increasing sizes: at each query, tuples are partitioned into groups, each group is loaded into a buffer through a kernel system call and is then read by routine {\tt mysql\_select}. The \rms\ does not count repeated buffer read operations: hence, the input size on larger tables is exactly the same as in smaller ones (it roughly coincides with the buffer size), while the running time grows due to the larger number of buffer loads.

Routine {\tt im\_generate} in benchmark {\tt vips} shows an analogous effect (see Figure~\ref{fig:vips-im-generate}). In this case the induced first-accesses not counted in the \rms\ are due to  the interaction between threads via shared memory. In both examples, the \rms\ plot appears to reveal an asymptotic bottleneck, which instead does not actually exist. In other cases, the scenario might be the opposite: the \rms\ may not reveal the existence of a possible performance bottleneck, which can be instead characterized using the \trms. For instance, the \trms\ plot of routine {\tt buf\_flush\_buffered\_writes} of {\tt MySQL} in Figure~\ref{fig:mysql-buf-flush} shows a superlinear running time, while the \rms\ plot  only suggests a linear growth, as highlighted by standard curve fitting techniques.

%--------------------------------------------------------------------------------
\paragraph{Profile richness.}

The usefulness of input-sensitive profile data crucially depends on the number of distinct input size values collected for each routine: each value corresponds to a point in the cost plots associated to a routine, and plots with a small number of points do not clearly expose the behavior of the routine. In our experiments, we observed that the use of \trms\ instead of \rms\ can often yield a larger number of distinct input size values and thus more informative plots.

An example is provided by Figure~\ref{fig:vips-wbuffer-write-thread}: while routine {\tt wbuffer\_write\_thread} was called 110 times during the execution of application {\tt vips}, according to the \rms\ metric all its input sizes collapsed onto two distinct values (67 and 69, as shown in Figure~\ref{fig:vips-wbuffer-write-thread}a). However, this routine performs many load operations from secondary memory: hence, if we also take into account external input (Figure~\ref{fig:vips-wbuffer-write-thread}b), or external input combined with thread-input (Figure~\ref{fig:vips-wbuffer-write-thread}c), the number of distinct \trms\ values grows considerably and the trend in the cost plots becomes more meaningful.

\begin{figure}[t]
\centering
\begin{tabular}{cc}
\hspace{-3mm}
\includegraphics[width=0.445\columnwidth]{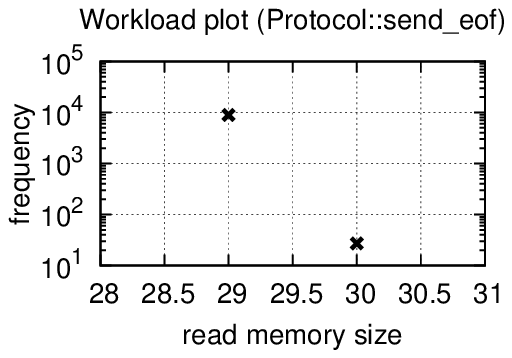} &
\includegraphics[width=0.445\columnwidth]{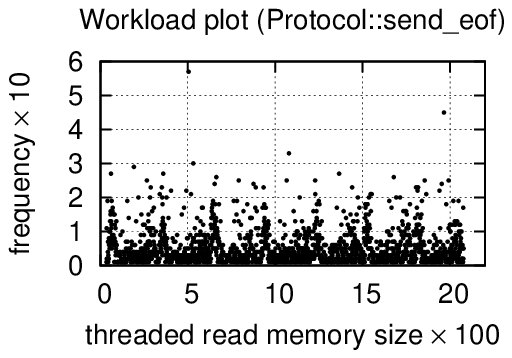}\hspace{-2mm} \\
\hspace{2.0mm}\small{(a)} & \hspace{4mm}\small{(b)} 
\end{tabular}
\caption{Function {\tt Protocal::send\_eof} of {\tt MySQL}: workload plots respectively obtained using
\rms\ or \trms\ as an estimate for the input size.}
\label{fig:mysql-send-eof}
\end{figure}

%--------------------------------------------------------------------------------
\paragraph{Workload and input characterization.}

An additional benefit of input-sensitive profiling is the capability of characterizing the typical workloads on which a routine is called in the context of deployed systems. Richer profile data collected using the \trms\ metric yield more accurate workload characterization, as shown, e.g., by the workload plots of Figure~\ref{fig:mysql-send-eof}. Moreover, our multithreaded input-sensitive profiling methodology can also provide insights on the amount of interaction of each routine with external devices (external input) and cooperating threads (thread-induced input): for instance,  $99.9\%$ of the input of routine {\tt wbuffer\_write\_thread} (Figure~\ref{fig:vips-wbuffer-write-thread}) is due to loads from secondary memory and thread interaction. 

\begin{figure*}[t]
\centering
\begin{tabular}{cc}
%\hspace{-3mm}
\includegraphics[width=0.95\columnwidth]{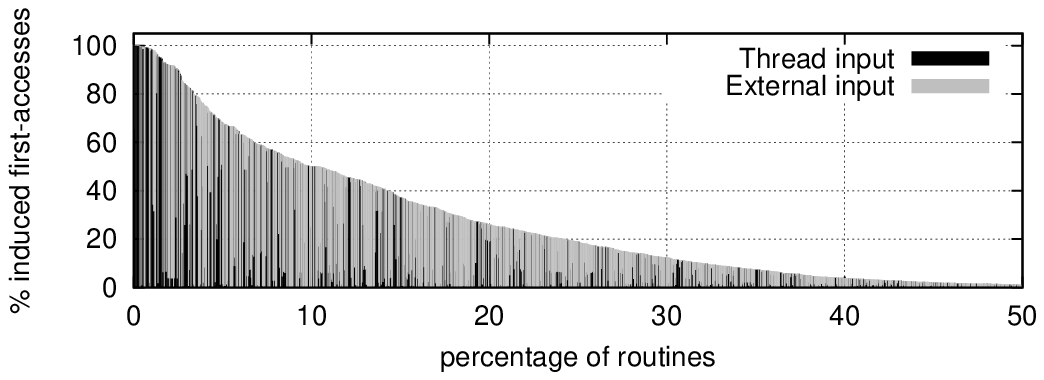}\hspace{+5mm} & \includegraphics[width=0.95\columnwidth]{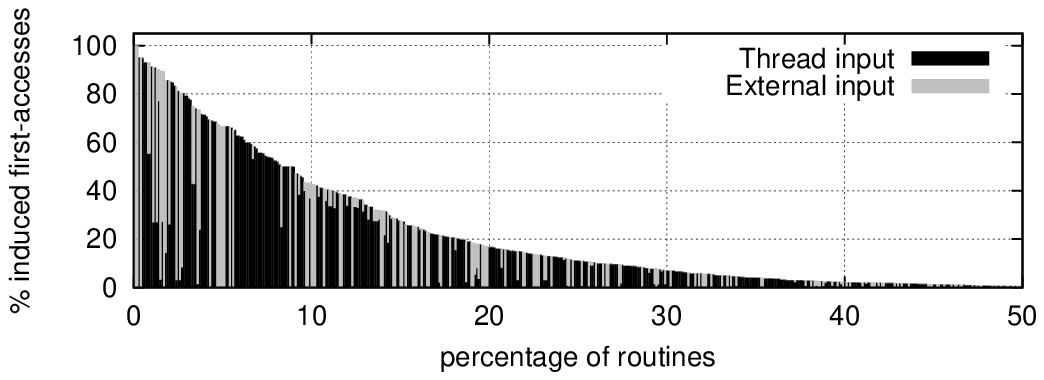} \\
\hspace{1mm}{\small{(a)}} & \hspace{6mm}{\small{(b)}}
%\hspace{4mm}\small{(a)} & \hspace{0mm}\small{(b)} & \hspace{0mm}\small{(c)} 
\end{tabular}
\caption{Thread-induced vs. external input on benchmarks (a) {\tt MySQL} and (b) {\tt vips}.}
\label{fig:thread-vs-external}
\end{figure*}

For each routine, we can automatically distinguish between external and thread-induced input. If we sort in decreasing order all routines in accordance with their percentage of induced first-accesses, we obtain an interesting characterization of the interplay between workload, computation, and concurrency, as shown in Figure~\ref{fig:thread-vs-external}. This figure plots, for each routine of benchmarks {\tt MySQL} and {\tt vips}, the percentage of induced first-accesses partitioned between external input and thread-induced input: a first look reveals that induced first-accesses of the majority of {\tt MySQL} routines are due to external input, differently from {\tt vips} routines where thread input is predominant. We remark that charts of this kind can be automatically produced by our profiler.
In Section~\ref{se:experiments} we will provide a quantitative evaluation of profile richness and input characterization in a variety of applications on typical workloads.

%Among induced first-accesses, we can also automatically identify how many accesses are due to external input or are thread-induced, obtaining graphs similar to those in Figure~\ref{fig:thread-vs-external}: this provides an interesting characterization of the interplay between workload, computation, and concurrency. Figure~\ref{fig:thread-vs-external} plots, for each routine of benchmarks {\tt MySQL} and {\tt vips}, the percentage of induced first-accesses partitioned between external input and thread-induced input. A first look reveals that induced fist-accesses in the majority of {\tt MySQL} routines are due to external input, differently from {\tt vips} routines where thread input is predominant. 
%
%We remark that graphs of this kind can be automatically produced by {\tt aprof-trms}.

\vspace{-1mm}
\section{Computing the Multithreaded Input Size}
\label{se:algorithms}

In this section we describe an efficient algorithm for computing the threaded read memory size of a routine activation and the input-sensitive profile of a routine. Routine profiles are thread-sensitive, i.e., profiles generated by routine activations made by different threads are kept distinct (if necessary, they can be combined in a subsequent step). 

The profiler is given as input multiple traces of program operations associated with timing information. Each trace is generated by a different thread and includes: routine activations ({\stt call}), routine completions ({\stt return}), {\stt read}/{\stt write} memory accesses, and read/write operations performed through kernel system calls ({\stt kernelRead} and {\stt kernelWrite}, necessary to characterize external input). 

As a first step, thread-specific traces are logically merged, interleaving operations performed by different threads according to their timestamps, in order to produce a unique execution trace. If two or more operations issued by different threads have the same timestamp, ties are broken arbitrarily: no assumption can be therefore done about which operation will be processed first. We remark that after merge and tie breaking, trace events are totally ordered. For simplicity of exposition, we also assume that {\stt switchThread} events are inserted in the merged trace between any two operations performed by different threads.

For each operation issued by a routine $r$ in a thread $t$, the profiler must update \trms\ and cost information of $r$ with respect to $t$. Some operations might also require to update profiling data structures related to threads other than $t$. To clarify the relationships between different threads, we first discuss a naive approach as a warm-up for the reader.

\vspace{-1mm}

%-----------------------------------------------------------
\subsection{Naive Approach}

%The activation of any routine $r$ by thread $t$ has its own \trms, which is part of the set $N_{r,t}$ of input sizes on which $r$ is invoked by $t$. 

Let $t$ be a thread and let $r$ be a routine activated by $t$. With a slight abuse of notation, we will denote with $\mbox{\trms}_{r,t}$ the threaded memory size of a specific activation of  $r$ in $t$. According to the definition of multithreaded input size (see Section~\ref{se:rvms}), computing $\mbox{\trms}_{r,t}$ requires to count read operations issued by routine $r$ that are either first accesses or induced first-accesses. In turn, identifying induced first-accesses requires to monitor write operations performed by {\em all} threads, i.e.,  performed also by threads different from $t$.

A simple-minded approach, which is sketched in Figure~\ref{fig:naive}, is to maintain a set $L_{r,t}$ of memory locations accessed during the activation of $r$. Immediately after entering $r$, this set is empty and $\mbox{\trms}_{r,t}=0$. Memory locations can be both added to and removed from $L_{r,t}$ during the execution of $r$, as follows: 
\begin{itemize}
\item when $r$ reads or writes a location $\ell$, then $\ell$ is added to $L_{r,t}$ (if not already present);

\item when a thread $t'\neq t$ writes a location $\ell$, then $\ell$ is removed from $L_{r,t}$ (if present): this allows it to recognize induced first-accesses. 

\end{itemize}

\begin{figure}[t]
\begin{tabular}{ |l | l| }
\hline
  $\mbox{\stt read}_t(\ell)$ & if $\ell\not\in L_{r,t}$ then $\mbox{\trms}_{r,t}$++  \\
    & ~~~~~~~~~~~~~~~~~~~~~~~~~$L_{r,t}\leftarrow L_{r,t}\cup\{\ell\}$  \\
  $\mbox{\stt write}_t(\ell)$ & $L_{r,t}\leftarrow L_{r,t}\cup\{\ell\}$  \\
  $\mbox{\stt read}_{t'}(\ell)$, ${t'}\neq t$ & --  \\
  $\mbox{\stt write}_{t'}(\ell)$, ${t'}\neq t$ & $L_{r,t}\leftarrow L_{r,t}\setminus\{\ell\}$\\\hline
\end{tabular}
\caption{Computation of $\mbox{\trms}_{r,t}$ with a naive approach. The notation $\mbox{\stt read}_t/\mbox{\stt write}_t(\ell)$indicates that location $\ell$ is read/written by thread $t$.}
\label{fig:naive}
\end{figure}

\noindent With this approach, at any time during the execution of $r$, a read operation on a location $\ell$ is a first access (possibly induced by other threads) if and only if $\ell\not\in L_{r,t}$. Hence, $\mbox{\trms}_{r,t}$ is increased only if this test succeeds. Notice that read operations performed by threads different from $t$ change neither set $L_{r,t}$ nor $\mbox{\trms}_{r,t}$.

We remark that in the description above $r$ can be any routine in the call stack of thread $t$ (not necessarily the topmost). Hence, the same checks and updates must be performed for all pending routine activations in the call stack of  $t$. Due to stack-walking and to the fact that write operations require to update sets $L_{r,t}$ of 
{\em all} threads, this simple-minded approach is extremely time-consuming. It is also very space demanding: in the worst case, each distinct memory location could be stored in all sets $L_{r,t}$ for each thread $t$ and each routine activation $r$ pending in the call stack of $t$. In that case, the space would be proportional to the memory size times the maximum stack depth times the number of threads.

%Stack walking dal topmost quando si inserisce (if we start from the topmost activation, it is possible to stop stack-walking when the first routine $r'$ is encountered such that $\ell$ already belongs to $L_{r',t}$), stack walking dal bottom quando si toglie

%-----------------------------------------------------------
\subsection{The Read/Write Timestamping Algorithm}
\label{ss:multi-threading}

To obtain a more space- and time-efficient algorithm, we exploit the latest-access approach described in~\cite{CDF12}.  Namely, we avoid to store explicitly the threaded read memory size $\mbox{\trms}_{r,t}$ and the sets $L_{r,t}$ of accessed memory locations. Instead, we maintain partial information that can be updated quickly during the computation and from which the \trms\ can be easily derived upon the termination of a routine. 

In more details, we adapt the latest-access approach~\cite{CDF12} as follows: for each thread $t$ and memory location $\ell$, we store $\ell$  in only one set $P_{r,t}$ such that $r$ is the latest routine activation in $t$ that accessed $\ell$ (either directly or by its completed subroutines). At any time during the execution of thread $t$ and for each pending routine activation $r$, it holds:
$$L_{r,t}=P_{r,t}\cup\{P_{r',t}:~r'~descendant~of~r\}$$
where $r'$ is any pending routine activation that is above $r$ in the call stack at that time. 
%
%At any time during the execution of thread $t$, the sets $P_{r,t}$ partition the locations accessed so far in $t$. 
%
Sets $P_{r,t}$ will be stored implicitly by associating timestamps to routines and memory locations. 

Similarly to the naive approach of Figure~\ref{fig:naive}, locations will be both added to and removed from $P_{r,t}$ to characterize induced first-accesses. However, this turns out to be inefficient in a multithreaded scenario: differently from read operations that change only thread-specific sets, write accesses require to change the sets $P_{r,t}$ of each activation $r$ pending in the call stack of each running thread $t$. By implicitly updating only one set $P_{r,t}$ per thread, the latest-access algorithm avoids stack walking, but the update time for write accesses is still linear in the number of threads, which can be prohibitive in practice. 

To overcome this problem, we combine the latest access approach with global timestamps that are appropriately updated upon write accesses to memory locations: in this way, we will recognize induced first-accesses by comparing thread-specific timestamps with global ones. The entire algorithm is sketched in Figure~\ref{fig:shadowing_algorithm}. 

\paragraph{Data structures.} 
The algorithm uses the following global data structures:
\begin{itemize}

\item a counter $count$ that maintains the total number of thread switches and routine activations across all threads;

\item a shadow memory $wts$ such that, for each memory location $\ell$, $wts[\ell]$ is the timestamp of the latest write operation on $\ell$ performed by any thread. The timestamp of a memory access is defined as the value of $count$ at the time in which the access took place.

\end{itemize}

\noindent Similarly to~\cite{CDF12}, the algorithm also uses the following thread-specific data structures for each thread $t$:
\begin{itemize}

\item a shadow memory $ts_t$ such that, for each memory location $\ell$, $ts_t[\ell]$ is the timestamp of the latest access (read or write) to $\ell$ made by thread $t$;

\item a shadow run-time stack $S_t$, whose top is indexed by variable $top_t$. For each $i\in [1,top_t]$, the $i$-{th} stack entry $S_t[i]$ stores:
\begin{itemize}
\item The id $S_t[i].rtn$, the timestamp $S_t[i].ts$, and the cumulative cost $S_t[i].\mbox{\em cost}$ of the $i$-th pending routine activation.
\item The {\em partial threaded read memory size} $S_t[i].trms$ of the activation, defined so that the following invariant property holds throughout the execution for each $i$ such that $1\le i\le top_t$:
\vspace{-3mm}
\begin{equation}
\label{invariant}
\hspace{-3mm}
\forall i,~1\le i\le top_t:~~
\mbox{\trms}_{i,t}=\sum_{j=i}^{top_t}S_t[j].trms
\vspace{-1.5mm}
\end{equation}
where $\mbox{\trms}_{i,t}$ is a shortcut for $\mbox{\trms}_{S_t[i].rtn,t}$. At any time, $\mbox{\trms}_{i,t}$ equals the current \trms\ value of the $i$-th pending activation on the portion of the execution trace generated by thread $t$ seen so far.
\end{itemize}

\end{itemize}

\begin{figure}[t]
\centering
%--------------------------------------
\begin{minipage}[t]{0.50\linewidth}
\flushleft
% ------- call -------
\textbf{procedure} {\tt call}$(r,t)$
\vspace{-.5cm}
\begin{algorithmic}[1]
\STATE $ count\,$++
\STATE $top_t\,$++
\STATE $S_t[top_t].rtn \leftarrow r$
\STATE $S_t[top_t].ts \leftarrow count$
\STATE $S_t[top_t].trms \leftarrow 0$
\STATE $S_t[top_t].cost \leftarrow $ \\
~~~~~~~~~~~~\verb+getCost()+
\end{algorithmic}
\vspace{.3cm}
% ------- return -------
\textbf{procedure} {\tt return}$(t)$
\vspace{-.5cm}
\begin{algorithmic}[1]
\STATE \verb+collect(+$S_t[top_t].rtn$,\\  
	~~~~~~~~~~~~$S_t[top_t].trms$,\\ 
	~~~~~~~~~~~~\verb+getCost()+-\\
	~~~~~~~~~~~~$S_t[top_t].cost$\verb+)+
\STATE $S_t[top_t\verb+-+1].trms\,$+=\\
~~~~~~~~~~~~$S_t[top_t].trms$
\STATE $top_t$\verb+--+
\end{algorithmic}
% ------- switch_thread -------
\vspace{.3cm}
\textbf{procedure} {\tt switchThread}$()$
\vspace{-.5cm}
\begin{algorithmic}[1]
\STATE{$count\,$++}
\end{algorithmic}
\end{minipage}
%--------------------------------------
\begin{minipage}[t]{0.49\linewidth}\flushleft
% ------- read -------
\textbf{procedure} {\tt read}$(\ell,t)$
\vspace{-.5cm}
\begin{algorithmic}[1]
\IF {$ts_t[\ell] < wts[\ell]$}
\STATE $S_t[top_t].trms\,$++
\ELSE 
\IF{$ts_t[\ell] < S_t[top_t].ts$}
\STATE $S_t[top_t].trms\,$++
\IF{$ts_t[\ell] \not= 0$}
\STATE \hspace{-1.7mm}$i=$ max idx s.t.\\
~$S_t[i].ts \leq ts_t[\ell]$
\STATE \hspace{-1.7mm}$S_t[i].trms\,$\verb+--+
\ENDIF 
\ENDIF
\ENDIF
\STATE $ts_t[\ell] \leftarrow count$
\end{algorithmic}
\vspace{.3cm}
% ------- write -------
\textbf{procedure} {\tt write}$(\ell,t)$
\vspace{-.5cm}
\begin{algorithmic}[1]
\STATE{ $ts_t[\ell] \leftarrow count$}
\STATE{$wts[\ell] \leftarrow count$}
\end{algorithmic}
\end{minipage}
\vspace{.1cm}
\caption{\trms\ algorithm: multi-threaded input.}
\label{fig:shadowing_algorithm}
\end{figure} 

\noindent As shown in~\cite{CDF12}, Invariant~\ref{invariant} implies the following interesting property: for each pending routine activation, its \trms\ value can be obtained by summing up its partial threaded read memory size with the \trms\ value  of its (unique) pending child, if any. More formally: $$\mbox{\trms}_{top_t,t}=S_t[top_t].trms$$
$$\mbox{\trms}_{i,t}=S_t[i].trms+\mbox{\trms}_{i+1,t}$$
for each $i\in[1,top_t-1]$. Hence, if we can correctly maintain the partial threaded read memory size during the execution, upon completion of a routine we will also get the correct \trms\ value.

%========================================
\paragraph{Algorithm and analysis.} 
The partial threaded read memory size can be maintained as shown in Figure~\ref{fig:shadowing_algorithm}. We first notice that the global timestamp counter $count$ is increased at each thread switch and routine call, and its value is used to update  routine timestamps (line 4 of procedure {\tt call}), global memory timestamps (line 2 of procedure {\tt write}), and local memory timestamps (lines 1 and 12 of procedures {\tt write} and {\tt read}, respectively). Upon activation of a routine, procedure {\tt call}$(r,t)$ creates and initializes a new shadow stack entry for routine $r$ in $S_t$. When the routine activation terminates, its cost is collected and its partial \trms\ (which at this point coincides with the true \trms\ value according to equation $\mbox{\trms}_{top_t,t}=S_t[top_t].trms$ discussed above) is added to the partial \trms\ of its parent, preserving Invariant~\ref{invariant}.

Local timestamps of memory locations are updated both by read and write accesses, while global timestamps are not updated upon read operations (they are thus associated to write operations only). This update scheme makes it possible to recognize induced first-accesses to any location $\ell$, which is done by lines 1-2 of procedure {\tt read}. If the  read/write timestamp $ts_t[\ell]$ local to thread $t$ is smaller than the global write timestamp $wts[\ell]$, then location $\ell$ must have been written more recently than the last read/write access to $\ell$ by thread $t$. Note that, if the latest access to $\ell$ was a write operation by thread $t$, then it would be  $ts_t[\ell]=wts[\ell]$ (see procedure {\tt write}), letting the test $ts_t[\ell]<wts[\ell]$ fail. Hence, if the test succeeds, the last write on $\ell$ must have been done by some thread $t'\neq t$, the read access by $t$ is an induced access, and the partial \trms\ of the topmost routine is correctly increased by line 2 of procedure {\tt read}. Invariant~\ref{invariant} is fully preserved by this assignment: the accessed value is new not only for the topmost routine in the call stack $S_t$, but also for all its ancestors, whose \trms\ is implicitly updated according to Equation~\ref{invariant}. 

On the other side, if the test of line 1 of procedure {\tt read} fails, the read access to $\ell$ might still be a first access: this happens if the last access to location $\ell$ by thread $t$ took place before entering the current (topmost) routine. Lines 4--10 address this case, updating the partial \trms\ as described in~\cite{CDF12}: the partial \trms\ of the topmost routine is increased, while the partial \trms\ of an appropariately chosen ancestor is decreased (it is proved in~\cite{CDF12} that Invariant~\ref{invariant} is preserved).

The running time of all operations is constant, except for line 7 of procedure {\tt read} that requires $O(\log d_t)$ worst case time, where $d_t$ is the depth of the call stack $S_t$.

%-----------------------------------------------------------
\subsection{External Input}
\label{ss:externalInput}

In Section~\ref{ss:multi-threading} we have focused on induced first-accesses generated by multi-threaded executions. In this section we show that the read/write timestamping algorithm can be naturally extended to take into account also induced accesses due to external input. 

Procedures {\tt kernelRead} and {\tt kernelWrite} shown in Figure~\ref{fig:shadowing_algorithm_2} update the profiler's data structures when memory accesses are mediated by kernel system calls. Threads invoke system calls to get data from external devices (e.g., a disk or the network) or to send data to external devices. We remark that the operating system kernel must be treated differently from normal threads in our algorithm, since there are no kernel-specific shadow memory and shadow stack.

\begin{figure}[t]
\centering
%---------------------------------------------
\begin{minipage}[t]{0.495\linewidth}\flushleft
% ------- input_on -------
\vspace{.3cm}
\textbf{procedure} {\tt kernelWrite}$(\ell)$
\vspace{-.5cm}
\begin{algorithmic}[1]
\STATE{{$count\,$++}}
\STATE{{$wts[\ell] \leftarrow count$}}
%\STATE{$wts[\ell] \leftarrow count$}
%\STATE{$ts_t[\ell] \leftarrow 0$}
\end{algorithmic}
\end{minipage}
%---------------------------------------------
\begin{minipage}[t]{0.495\linewidth}\flushleft
% ------- output_from -------
\vspace{.3cm}
\textbf{procedure} {\tt kernelRead}$(\ell,t)$
\vspace{-0.95cm}
\begin{algorithmic}[1]
\STATE{{\tt read}$(\ell,t)$}
\end{algorithmic}
\end{minipage}
\vspace{.1cm}
\caption{\trms\ algorithm: external input.}
\label{fig:shadowing_algorithm_2}
\end{figure} 

When a thread sends data to an external device, it must delegate the operating system to read the memory locations containing those data and write their content to the device. Hence, a thread external write operation corresponds to a {\tt kernelRead} event in the execution trace. As shown in Figure~\ref{fig:shadowing_algorithm_2}, read memory accesses by the operating system are regarded as read operations implicitly performed by the thread, as if the system call were a normal subroutine. Upon a  {\tt kernelRead} event, it is therefore sufficient to invoke procedure {\tt read} that, if necessary, will update the threaded \trms\ of pending routine activations. 

The case of {\tt kernelWrite} operations is slightly more subtle. When a thread needs data from a external device, it delegates the operating system to write the device data to some memory buffer (if the buffer consists of $n$ memory locations, the execution trace will contain $n$ distinct {\tt kernelWrite} events). This buffer load, however, should not be regarded as a thread external read operation: indeed, it may happen that only a subset of the loaded memory locations will be actually processed (and thus read) by the thread, and only those subset should be counted as external input. For this reason procedure {\tt kernelWrite} does not directly change the partial \trms\ of the topmost routine. Instead, it first increases $count$ and then associates buffer memory locations with a global write timestamp that is larger than any thread-specific timestamp. This forces the test $ts_t[\ell]<wts[\ell]$ to succeed if a buffer location $\ell$ will be subsequently read by the thread, properly increasing the partial \trms\ only for actual read operations.

%-----------------------------------------------------------
\subsection{Counter Overflows}

The global counter used by the read/write timestamping algorithm is common to all running threads and in our initial experiments was affected by frequent overflows, especially for long-running applications. Unfortunately, overflows are a serious concern in the computation of the \trms, since they alter the partial ordering between memory timestamps yielding wrong input size values. To overcome this issue, we perform a periodical global renumbering of timestamps in the profiler's data structures, taking care not to alter the partial order between $ts_t[\ell]$, $wts[\ell]$, and $S_t[i].ts$ for each memory location $\ell$, running thread $t$, and $1\le i\le top_t$. Instead, we exploit the following observation: in Figure~\ref{fig:shadowing_algorithm} there is no comparison between $wts[\ell]$ and $wts[\ell']$ or between $ts_t[\ell]$ and $ts_t[\ell']$, for $\ell\neq\ell'$. Hence, the order between timestamps of different memory locations is irrelevant and can be arbitrarily changed. 

Our renumbering algorithm is sketched in Figure~\ref{fig:overflow_algorithm}. For the sake of efficiency, the algorithm checks and renumbers each timestamp only once. Lines 1-4 collect all timestamps in the call stacks of running threads and sort them in increasing order. Notice that these timestamps are distinct: $count$ is  increased by procedure {\tt call} (see Figure~\ref{fig:shadowing_algorithm}) so that a new activation is always assigned an unused value, and the renumbering algorithm keeps the property true. 

Routine timestamps are re-assigned in lines 5-8: the new timestamps are multiples of 3 (this choice will be justified below) and are chosen according to the rank of the original routine timestamp in the sorted set $A$. This guarantees that the original ordering between any two routine timestamps is preserved, and that the maximum value of a timestamp will be proportional to the total number of pending routine activations (i.e., $|A|$).

\begin{figure}[t]
\centering
%-----------------------------------------------
\begin{minipage}[t]{0.975\linewidth}
\flushleft
% ------- call -------
\textbf{procedure} \verb+counterOverflow()+
\vspace{-.5cm}
\begin{algorithmic}[1]
\STATE \textbf{for each} {running thread $t$} \textbf{do}
\STATE ~~~~\textbf{for} $i = 1$ \textbf{to} $top_t$ \textbf{do}
\STATE ~~~~~~~~add $S_t[i].ts$ to set $A$ of active timestamps
\STATE {\tt sort}$(A)$
\STATE \textbf{for each} {running thread $t$} \textbf{do}
  \STATE ~~~~\textbf{for} $i = 1$ \textbf{to} $top_t$ \textbf{do} 
  \STATE ~~~~~~~~$p=$ position of timestamp $S_t[i].ts$ in $A$
  \STATE ~~~~~~~~$S_t[i].ts = 3 \cdot p$
\STATE \textbf{for each} {memory location $\ell$} \textbf{do}
  \STATE ~~~~$q=$ max idx in $A$ s.t. $wts[\ell] \leq A[q]$
  \STATE ~~~~\textbf{for each} {running thread $t$} \textbf{do}
    \STATE~~~~~~~~\textbf{if}~$ts_t[\ell] < A[q] \,\vee\, ts_t[\ell] \geq A[q+1]$~\textbf{then}
        \STATE~~~~~~~~~~~~$j=$ max idx in $A$ s.t. $ts_t[\ell] \geq A[j]$
        \STATE~~~~~~~~~~~~$ts_t[\ell] = 3 \cdot j$
    \STATE ~~~~~~~~\textbf{elif} $wts[\ell] = ts_t[\ell]$ \textbf{then} $ts_t[\ell] = 3 \cdot q + 1$
    \STATE ~~~~~~~~\textbf{elif} $wts[\ell] > ts_t[\ell]$ \textbf{then} $ts_t[\ell] = 3 \cdot q$
    \STATE ~~~~~~~~\textbf{else} $ts_t[\ell] = 3 \cdot q + 2$
%    \STATE \textbf{end if}
%  \ENDFOR
  \STATE ~~~~$wts[\ell] = 3 \cdot q + 1$
  \STATE $count = 3 \cdot |A| + 3$
%\ENDFOR
\end{algorithmic}
\end{minipage}
\vspace{.1cm}
\caption{Counter overflow procedure.}
\label{fig:overflow_algorithm}
\end{figure} 

Thread-specific and global timestamps of memory locations are re-assigned in lines 9-18. According to line 10, let $A[q]$ be the latest pending routine activation (in any thread) started before the latest write to memory location $\ell$. A thread $t$ could have accessed $\ell$ before this activation (i.e., $ts_t[\ell] < A[q]$), between pending routine activations $A[q]$ and $A[q+1]$ (i.e., $A[q] \leq ts_t[\ell] < A[q+1]$), or after pending routine activation $A[q+1]$ (i.e., $ts_t[\ell] \geq A[q+1]$). If $q+1$ is not a valid index for $A$ and $ts_t[\ell] \geq A[q]$ we can assume to be in the second case. The first and the third cases can be treated by assigning $ts_t[\ell]$ with the same value used for the most recent activation $j$ such that $ts_t[\ell] \geq A[j]$ (lines 12-14): this guarantees that comparisons between $ts_t[\ell]$ and any routine timestamp at lines 4 and 7 of procedure {\tt read} will succeed if and only if they succeeded before renumbering. The second case ($A[q] \leq ts_t[\ell] < A[q+1]$) requires to distinguish between three different situations,  which explains why new routine timestamps are chosen in line 8 as multiples of 3:
\begin{enumerate}
    
    \item $wts[\ell]=ts_t[\ell]$: $t$ was the last thread to write  location $\ell$. After renumbering (lines 15 and 18), $ts_t[\ell]=wts[\ell]=3q+1$. This guarantees that both $A[q]=3q\leq ts_t[\ell]=wts[\ell]=3q+1< A[q+1]= 3(q+1)$;
    
    \item $wts[\ell] > ts_t[\ell]$: thread $t$ has accessed location $\ell$ before its last write. In this case the new timestamp of $ts_t[\ell]$ is $3q$ (line 16). This preserves both the relations $ts_t[\ell]=3q<wts[\ell]=3q+1$ and  $ts_t[\ell]=3q<A[q+1]=3(q+1)$; 
    
    \item $ts_t[\ell] > wts[\ell]$: thread $t$ has read location $\ell$ after its last write. In this case $ts_t[\ell]$ gets the new value $3q + 2$ (line 17).  The order relation $wts[\ell]=3q+1 < ts_t[\ell]=3q+2< A[q+1]=3(q+1)$  remains valid.
\end{enumerate}
\noindent Notice that the global timestamp $count$ is assigned with a value larger than all the other timestamps (line 19).

Using binary search to implement lines 7, 10, and 13, the running time of the global renumbering algorithm is $O(\rho \log\rho + \mu\,\tau\log\rho)$ where $\tau$, $\mu$, and $\rho$ are the numbers of running threads, distinct memory cells, and pending routine activations, respectively. This can be amortized against $\Omega(2^w)$ thread switch and routine call operations, where $w$ is the word memory size.

% !TEX root = paper.tex
%--------------------------------------------------------------------------------
\section{Implementation}
\label{se:implementation}

%In this section we discuss our implementation of {\tt aprof-trms} based on the Valgrind~\cite{NethercoteS07} framework.
To prove the feasibility of our approach, we implemented a threaded input-sensitive profiler by developing a Valgrind~\cite{NethercoteS07} tool called {\tt aprof-trms}.
%In this section we provide some details on our implementation of \aprof, a Valgrind tool for input-sensitive profiling. 
Valgrind provides a dynamic instrumentation infrastructure that translates the binary code into an architecture-neutral intermediate representation (VEX). Analysis tools provide callbacks for events generated by the stream of VEX executed instructions. 

\paragraph{Instrumentation.} Similarly to the input-sensitive profiler described in~\cite{CDF12}, our tool traces all memory accesses and function calls and returns. We count basic blocks as a performance measure: tracing function calls and returns requires to instrument each basic block, thus counting basic blocks adds a light burden to the analysis time overhead, and improves accuracy in characterizing asymptotic behavior even on small workloads. Measuring basic blocks rather than time has several other advantages, very well motivated in~\cite{GAW07}. In order to take into account external input, system calls are wrapped and properly mapped to one or more {\tt kernelRead} or {\tt kernelWrite} events:  among the main system calls on a Linux x86\_64 machine, {\tt write}, {\tt sendto}, {\tt pwrite64}, {\tt writev}, {\tt msgsnd}, and {\tt pwritev} correspond to {\tt kernelRead} events, while {\tt read}, {\tt recvfrom}, {\tt pread64}, {\tt readv}, {\tt msgrcv}, and {\tt preadv} correspond to {\tt kernelWrite} events.

%Table~\ref{fig:syscall} provides a brief summary of this mapping for the main syscalls on a Linux x86\_64 machine. 

%\begin{figure}[t]
%\centering
%\begin{tabular}{l||c|c|}
%%\cline{2-3}
%\multicolumn{1}{c||}{} & {\tt kernelRead} & {\tt kernelWrite} \\ \hline
%
%{\tt write} & $\times$ & \\
%%{\tt send} & $\times$ & \\
%{\tt sendto} & $\times$ & \\
%{\tt pwrite64} & $\times$ & \\
%{\tt writev} & $\times$ & \\
%{\tt msgsnd} & $\times$ & \\
%{\tt pwritev} & $\times$ & \\
%\hline
%{\tt read} & & $\times$  \\
%{\tt recvfrom} & & $ \times$ \\
%{\tt pread64} & & $ \times$ \\
%{\tt readv} & & $ \times$ \\
%{\tt msgrcv} & & $ \times$ \\
%{\tt preadv} & & $ \times$ \\
%%{\tt mmap} & & $ \times$ \\
%\hline
%\end{tabular}
%\caption{Mapping between system calls and {\tt kernelRead} or {\tt kernelWrite} events on a Linux x86\_64 machine.}
%\label{fig:syscall}
%\end{figure}

\paragraph{Thread interleaving.} Under Valgrind, threads of a traced application are serialized. This makes the development and debugging of a dynamic analysis framework and of its derived tools easier. 
% tools built on top of it. 
Serialization should not be seen as a crucial limitation of our implementation: for instance Helgrind~\cite{MW07,ValgrindTools} and DRD~\cite{ValgrindTools}, two popular tools for detecting synchronization errors in programs that use the POSIX {\tt pthreads} primitives, are both based on Valgrind. Serialization implies that our profiler does not need to perform tie breaking of events (see Section~\ref{se:algorithms}). However, in a serialized scenario the  scheduling of threads becomes a critical concern: thread interleaving may be altered and the execution may deviate from non-serialized executions. In order to avoid unrealistic executions, our tool takes benefit of the fair thread scheduler introduced in the latest release of Valgrind. 

\paragraph{Shadow memories.} To reduce space overhead in practice, we maintain global and thread-specific shadow memories using three-levels lookup tables. A similar approach is also adopted by other prominent tools, such as {\tt memcheck}~\cite{SN05}. A primary table indexes 2048 secondary tables, each covering 1GB of address space by indexing 16K chunks. Each chunk, in turn, contains the set of 32-bit timestamps for 64KB address space. In this way only chunks related to memory cells actually accessed by a thread need to be shadowed in its thread-specific shadow memory. Hence, on average (e.g., with embarrassingly parallel applications), the accessed primary memory is roughly partitioned among all running threads: the overhead for maintaining global and thread-specific shadow memories is therefore considerably smaller than in the worst-case scenario (where it would be proportional to number of running threads $\times$ number of distinct memory cells). Experiments in Section~\ref{se:experiments} will largely confirm this hypothesis.

% !TEX root = paper.tex
%--------------------------------------------------------------------------------
\section{Experimental Evaluation}
\label{se:experiments}

\begin{table*}[t]
\begin{scriptsize}
\begin{center}
\begin{tabular}{|l||c|c|c|c|c|c|c||c|c|c|c|c|c|c|}\hline
\multicolumn{1}{|c||}{} &
\multicolumn{7}{|c||}{\sc Time} &
\multicolumn{7}{|c|}{\sc Space} \\\cline{2-15}

\multicolumn{1}{|c||}{} & 
\multicolumn{1}{|c|}{secs} &
\multicolumn{6}{|c||}{slowdown} &
\multicolumn{1}{|c|}{MB} &
\multicolumn{6}{|c|}{overhead} \\\cline{2-15}

\multicolumn{1}{|c||}{} & 

\multirow{2}{*}{native} &
\multicolumn{1}{|c|}{{\tt\scriptsize nul}} & 
\multicolumn{1}{|c|}{{\tt\scriptsize mem}} & 
\multicolumn{1}{|c|}{{\tt\scriptsize call}} & 
\multicolumn{1}{|c|}{{\tt\scriptsize hel}} & 
\multicolumn{1}{|c|}{{\tt\scriptsize aprof}} & 
\multicolumn{1}{|c||}{{\tt\scriptsize aprof}} & 

\multirow{2}{*}{native} &
\multicolumn{1}{|c|}{{\tt\scriptsize nul}} & 
\multicolumn{1}{|c|}{{\tt\scriptsize mem}} & 
\multicolumn{1}{|c|}{{\tt\scriptsize call}} & 
\multicolumn{1}{|c|}{{\tt\scriptsize hel}} & 
\multicolumn{1}{|c|}{{\tt\scriptsize aprof}}  &
\multicolumn{1}{|c|}{{\tt\scriptsize aprof}}  \\

\multicolumn{1}{|c||}{} &
 
\multicolumn{1}{|c|}{} &
\multicolumn{1}{|c|}{{\tt\scriptsize grind}} &
\multicolumn{1}{|c|}{{\tt\scriptsize check}} &
\multicolumn{1}{|c|}{{\tt\scriptsize grind}} &
\multicolumn{1}{|c|}{{\tt\scriptsize grind}} &
\multicolumn{1}{|c|}{{\tt\scriptsize rms}} &
\multicolumn{1}{|c||}{{\tt\scriptsize trms}} &

\multicolumn{1}{|c|}{} &
\multicolumn{1}{|c|}{{\tt\scriptsize grind}} &
\multicolumn{1}{|c|}{{\tt\scriptsize check}} &
\multicolumn{1}{|c|}{{\tt\scriptsize grind}} &
\multicolumn{1}{|c|}{{\tt\scriptsize grind}} &
\multicolumn{1}{|c|}{{\tt\scriptsize rms}} &
\multicolumn{1}{|c|}{{\tt\scriptsize trms}} \\\hline

\texttt{350.md}       & 3184.5 & 6.0   & 43.5  & 34.7  & 125.4 & 39.6  & 41.2  & 50.8   & 1.9 & 2.0 & 2.0 & 6.7  & 2.2 & 2.3 \\
\texttt{351.bwaves}   & 192.0  & 22.1  & --    & 68.4  & 92.0  & 78.0  & 91.2  & 1582.1 & 1.1 & --  & 1.1 & 4.0  & 3.0 & 4.0 \\
\texttt{352.nab}      & 185.0  & 21.1  & 111.4 & 80.4  & 127.2 & 107.5  & 186.7 & 57.0   & 1.7 & 2.4 & 1.8 & 5.8  & 2.5 & 2.8 \\
\texttt{358.botsalgn} & 8.0    & 42.8  & 85.6  & 132.9 & 114.2 & 146.3  & 179.3 & 57.7   & 1.6 & 1.7 & 1.7 & 3.4  & 1.8 & 2.3 \\
\texttt{359.botsspar} & 1.0    & 204.0 & 353.9 & 523.1 & 368.3 & 301.4  & 457.6 & 62.9   & 1.6 & 2.1 & 1.7 & 4.8  & 2.4 & 2.6 \\
\texttt{360.ilbdc}    & 1936.8 & 2.2   & 18.5  & 4.7   & 64.9  & 17.3  & 26.2  & 1415.2 & 1.0 & 1.1 & 1.0 & 1.8  & 4.1 & 5.1 \\
\texttt{362.fma3d}    & 46.6   & 22.7  & 113.4 & 45.2  & 393.4 & 99.0  & 118.0 & 155.8  & 1.3 & 2.5 & 1.3 & 10.6 & 3.0 & 3.8 \\
\texttt{367.imagick}  & 170.0  & 24.1  & 91.0  & 50.5  & 141.6 & 52.3  & 60.6  & 77.9   & 1.6 & 2.0 & 1.7 & 5.4  & 3.1 & 3.9 \\
\texttt{370.mgrid331} & 5.2    & 39.7  & 101.7 & 50.9  & 194.6 & 95.7  & 130.2 & 395.1  & 1.1 & 1.2 & 1.1 & 1.8  & 2.1 & 3.0 \\
\texttt{371.applu331} & 26.6   & 45.0  & 230.2 & 103.7 & 472.8 & 228.9  & 367.6 & 88.2   & 1.5 & 1.8 & 1.6 & 5.6  & 3.2 & 3.7 \\
\texttt{372.smithwa}  & 14.4   & 28.8  & 78.4  & 57.0  & 148.1 & 167.8  & 213.9 & 49.4   & 1.8 & 1.9 & 2.0 & 3.3  & 2.4 & 2.6 \\
\texttt{376.kdtree}   & 33.9   & 19.5  & 99.4  & 127.3 & 366.6 & 247.4 & 551.0 & 98.6   & 1.4 & 2.8 & 1.5 & 8.4  & 4.4 & 5.6 \\

\hline\hline
{geometric mean}      &        & 23.6  & 94.1  & 64.8  & 179.4 & 101.5 & 140.8 &        & 1.4 & 2.0 & 1.5 & 4.5  & 2.8 & 3.3 \\\hline

\end{tabular}

\end{center}
\end{scriptsize}
\caption{Performance comparison of {\tt aprof-trms} with {\tt aprof} and some prominent Valgrind tools on the SPEC OMP2012 benchmarks.}
\label{ta:perf-table}
\end{table*}

In this section we discuss the results of an extensive experimental evaluation of {\tt aprof-trms} on a variety of benchmarks, including the SPEC OMP2012~\cite{SPECOMP2012} and the Princeton Application Repository for Shared-Memory Computers (PARSEC 2.1)~\cite{BKSL08}. The goals of our experiments are threefold: studying the slowdown and space overhead of {\tt aprof-trms} compared to other heavyweight dynamic analysis tools, evaluating the benefits of the \trms\ with respect to the \rms, and characterizing the amount of thread-induced and external input on the considered benchmarks.

%-----------------------------------------------------------------------
\subsection{Experimental Setup} 

%-----------------------------------------------
\paragraph{Benchmarks.} The OMP2012 benchmark suite of the Standard Performance Evaluation Corporation~\cite{SPECOMP2012} is a collection of fourteen OpenMP-based applications from different science domains. All of them were run on the SPEC input {\tt train} workloads in 64-bit mode.  We could successfully test all the components except for {\tt bt331} and {\tt swim}, whose execution failed due to a Valgrind memory issue.

The Princeton Application Repository for Shared-Memory Computers (PARSEC 2.1) is a benchmark
suite for studies of Chip-Multiprocessors~\cite{BKSL08}. It includes different workloads chosen from a variety of areas such as computer vision, media processing, computational finance, enterprise servers, and animation physics. PARSEC defines six input sets for each benchmark: experimental results reported in this section are all based on the {\tt simlarge} sets~\cite{BKSL08}.

For the sake of completeness, we also included in our tests the {\tt MySQL} application (version 5.5.30) discussed in Section~\ref{se:case-studies}: we used the {\tt mysqlslap} load emulation client~\cite{mysqlslap}, simulating 50 concurrent clients that submit approximately 1000 auto-generated queries.

%-----------------------------------------------
\paragraph{Metrics.} Besides slowdown and space overhead, we use the following metrics:

\begin{enumerate}

\item {\em Routine profile richness}: for each routine $r$, let $|\mbox{\rms}_r|$ be the number of distinct \rms\ values collected for routine $r$ (each value corresponds to a point in the graphs associated with $r$). Similarly, let
$|\mbox{\trms}_r|$ be the number of distinct \trms\ values collected for routine $r$ for all threads. The {profile richness} of routine $r$ is defined as: 
$$\frac{|\mbox{\trms}_r|-|\mbox{\rms}_r|}{|\mbox{\rms}_r|}$$
Intuitively, this metric compares the number of distinct input values obtained using the \trms\ and the \rms, respectively. We notice that $|\mbox{\trms}_r|\geq|\mbox{\rms}_r|$ does not necessarily hold: it may happen that two distinct \rms\ values $x$ and $y$ (obtained from two different activations of a routine) correspond to the same \trms\ value $z$, with $z\ge \max\{x,y\}$. Hence, the profile richness may be either positive, if more points are collected using the \trms, or negative, if more points are collected using the \rms. We will see that in practice the latter case is seldom true.

\item {\em Input volume}: according to Inequality~\ref{eq:trms-larger}, the \trms\ of a routine activation is always larger than or equal to the \rms\ of the same activation. The input volume metric characterizes the increase of the input size values due to multi-threading and to external input for an entire execution:
$$1-\frac{\sum_{routine~activations~r} \mbox{\rms}_r}{\sum_{routine~activations~r} \mbox{\trms}_r}$$
Values of this metric range in $[0,1)$. If $\mbox{\trms}_r=\mbox{\rms}_r$ for all routine activations $r$, then the input volume is 0. Conversely, if $\mbox{\trms}_r\gg\mbox{\rms}_r$ for all routine activations $r$, then the input volume gets close to 1. 

\item {\em Thread-induced input}: this metric measures the percentage of induced first-accesses (line 2 of procedure {\tt read} in Figure~\ref{fig:shadowing_algorithm}) due to multi-threading.

\item {\em External input}: similarly to the previous case, this metric measures the percentage of induced first-accesses due to external input.

\end{enumerate}

%-----------------------------------------------
\paragraph{Evaluated tools.} 

We compared the performance of {\tt aprof-} {\tt trms} to four reference Valgrind tools: {\tt nulgrind},  which does not collect any useful information and is used only for testing purposes, {\tt memcheck}~\cite{SN05},  a tool for detecting memory-related errors,  {\tt callgrind}~\cite{WKT04}, a call-graph generating profiler, and {\tt helgrind}~\cite{MW07}, a data race detector.  Although the considered tools solve different analysis problems, all of them share the same instrumentation infrastructure provided by Valgrind, which accounts for a significant fraction of the execution times: {\tt memcheck} does not trace function calls/returns and mainly relies on memory read/write events; {\tt callgrind} instruments function calls/returns, but not memory accesses, and {\tt helgrind} analyzes concurrent applications. We also compared {\tt aprof-trms} against a previous version of {\tt aprof} based on the \rms\ metric (see Section~\ref{se:rvms}): we remark that {\tt aprof-rms} targets sequential computations only, without taking into account induced first-accesses.

%-----------------------------------------------
\paragraph{Platform.} Experiments were performed on a cluster machine with four nodes, each equipped with two 64-bit AMD Opteron Processors 6272 @ 2.10 GHz (32 cores), with 64 GB of RAM running Linux kernel 2.6.32 with {\tt gcc} 4.4.7 and Valgrind 3.8.1 -- SVN rev. 13126.

\begin{figure}[t]
\centering
\begin{tabular}{cc}
\hspace{-3mm}
\includegraphics[width=0.445\columnwidth]{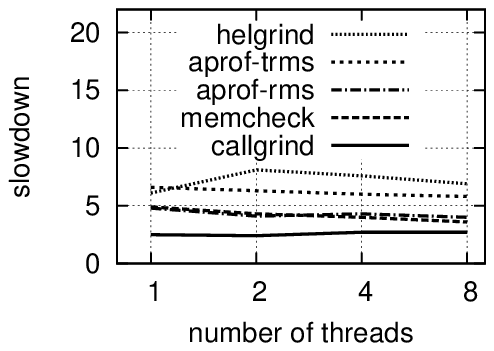} &
\includegraphics[width=0.445\columnwidth]{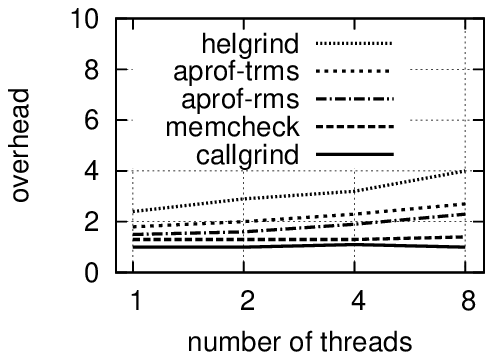}\hspace{-2mm} \\
\hspace{3mm}\small{(a)} & \hspace{7mm}\small{(b)} 
\end{tabular}
\caption{(a) Time and (b) space overhead, with respect to {\tt nulgrind}, as a function of the number of threads.}
\label{fig:threads}
\end{figure}

\begin{figure*}[t]
\centering
\begin{tabular}{cccccc}
\hspace{-3mm}
{\raisebox{13mm}{\small{(a)}}} & \includegraphics[width=0.445\columnwidth]{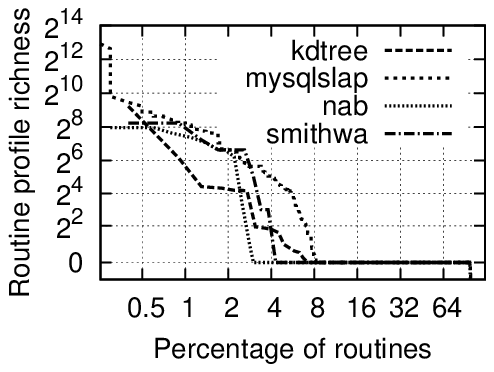}\hspace{+5mm} &
{\raisebox{13mm}{\small{(b)}}} & \includegraphics[width=0.445\columnwidth]{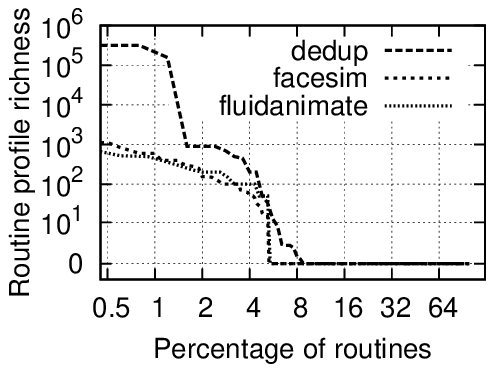}\hspace{+5mm} &
{\raisebox{13mm}{\small{(c)}}} & \includegraphics[width=0.445\columnwidth]{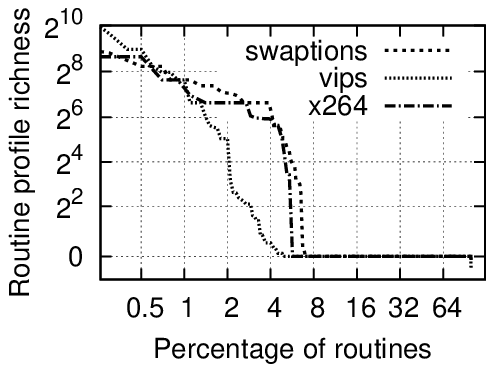}
%\hspace{4mm}\small{(a)} & \hspace{0mm}\small{(b)} & \hspace{0mm}\small{(c)} 
\end{tabular}
\caption{Routine profile richness of \trms\ w.r.t. \rms\ for a representative set of benchmarks.}
\label{fig:richness}
\end{figure*} 

\begin{figure*}[t]
\centering
\begin{tabular}{cccccc}
\hspace{-3mm}
{\raisebox{13mm}{\small{(a)}}} & \includegraphics[width=0.445\columnwidth]{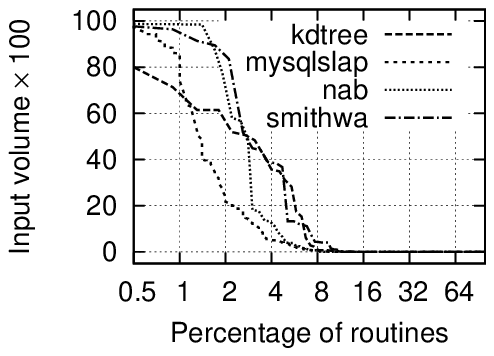}\hspace{+5mm} &
{\raisebox{13mm}{\small{(b)}}} & \includegraphics[width=0.445\columnwidth]{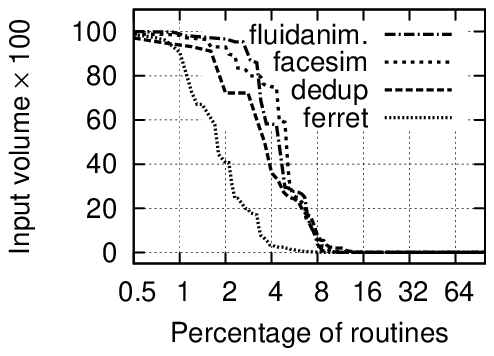}\hspace{+5mm} &
{\raisebox{13mm}{\small{(c)}}} & \includegraphics[width=0.445\columnwidth]{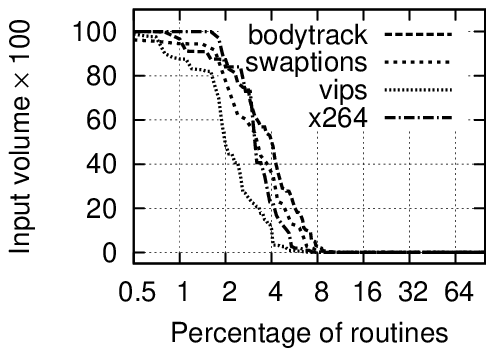}
%\hspace{4mm}\small{(a)} & \hspace{0mm}\small{(b)} & \hspace{0mm}\small{(c)} 
\end{tabular}
\caption{Input volume of \trms\ w.r.t. \rms\ for a representative set of benchmarks.}
\label{fig:input-volume}
\end{figure*}

%-----------------------------------------------------------------------
\subsection{Experimental Results}

%--------------------------------------------------------
\paragraph{Slowdown and space overhead.}

Performance figures of our evaluated tools on the SPEC OMP2012 benchmarks, obtained spawning four OpenMP threads per benchmark, are summarized in Table~\ref{ta:perf-table}. We do not report results for the PARSEC benchmarks because some tools revealed invalid memory accesses and other memory issues that prevented a reliable comparison of executions under different tools.

Compared to native execution, all the evaluated tools exhibit a large slowdown: even {\tt nulgrind}, which is reported to be roughly 5 times slower than native~\cite{ValgrindTools}, in our experiments  turned out to have  a mean slowdown factor of $23.6\times$. {\tt aprof-trms} is on average 6 times slower than {\tt nulgrind}: this is worse than {\tt memcheck}, which is $1.5$ times faster than our tool but does not trace function calls and returns, and better than {\tt helgrind}, which is $1.3$ times slower than {\tt aprof-trms} and is the only tool designed for the analysis of concurrent computations. 
Recognizing induced first-accesses causes a $38\%$ overhead on the running time, as demonstrated by the comparison of {\tt aprof-trms} with  {\tt aprof-rms}.

The mean memory requirements of {\tt aprof-trms} are within a factor of $3.3$ of native execution. This confirms our expectation: if memory is roughly partitioned among the four threads, the three-level lookup tables guarantee that the overall size of thread-specific shadow memories is proportional to the size of accessed memory locations. This should be added to the size of the global shadow memory, thus obtaining the total $3.3\times$ space overhead. Even tools that do not use shadowing, such as {\tt nulgrind} and {\tt callgrind}, require at least $1.4\times$ more space than native execution. {\tt memcheck} hinges upon memory shadowing, but turns out to be more efficient than {\tt aprof-trms} thanks to the adoption of memory compression schemes and to its independence from the number of threads. Similarly, {\tt aprof-rms} is slightly more efficient than our tool due to the lack of a global shadow memory. On the other hand we remark that {\tt helgrind}, which is akin to our tool with respect to the analysis of concurrency issues,  uses $36\%$ more space than {\tt aprof-trms}. 

Figure~\ref{fig:threads} shows the average slowdown and space overhead, with respect to the reference Valgrind tool {\tt nulgrind}, as a function of the number of spawned OpenMP threads. 
%In order to cut out the impact of Valgrind's thread serialization, results are computed with respect to {\tt nullgrind}. 
All the evaluated tools appear to scale properly. The average slowdown slightly decreases as the number of threads increases: this is because Valgrind serializes the execution of threads, and the time spent for instrumentation can be better amortized over the serialized executions of a larger number of threads. Overall, this experiment confirms the results detailed in Table~\ref{ta:perf-table} in the case of four threads. The mean space overhead of {\tt callgrind} and {\tt memcheck} is roughly constant: their analyses are indeed independent from concurrency issues. On the other hand, {\tt aprof-trms} and {\tt helgrind} show a modest increase of the space overhead when the number of threads increases: our profiling of the memory usage of {\tt aprof-trms} revealed that the space overhead  mostly depends on shadow memories, whose total space usage, however,  grows sublinearly with the number of threads. This confirms the effectiveness of our implementation based on three-level lookup tables. The comparison with {\tt aprof-rms} also suggests that the space overhead due to the global shadow memory used by {\tt aprof-trms} is better amortized as the number of threads increases.

%\vspace*{-2.0mm}
%--------------------------------------------------------
\paragraph{TRMS versus RMS.} Our second set of experiments aims at evaluating the benefits of the \trms\ metric with respect to the \rms. As shown in~\cite{CDF12}, an \rms-based input-sensitive profiler can collect a significant number of distinct input values for most algorithmic-intensive functions, thus producing informative cost plots. A first natural question is whether using \trms\ instead of \rms\ has any positive or negative impact on the profile richness. Charts in Figure~\ref{fig:richness} contribute to answer this question. Each curve is related to a specific benchmark. A point $(x,y)$ on a curve means that $x\%$ of routines have profile richness at least $y$: e.g., in benchmark {\tt dedup},  the number of points collected using the \trms\ metric is more than $100$ times larger than using the \rms\ for roughly $4\%$ of the routines.  
As expected, only for a small percentage of routines $|\mbox{\trms}_r|$  is much larger than $|\mbox{\rms}_r|$: this is due to the fact that I/O and thread communication are typically encapsulated in a small number of software components. However, for these routines $|\mbox{\trms}_r|$  can be substantially larger than $|\mbox{\rms}_r|$, e.g., up to a factor of roughly $10^6$ for benchmark {\tt dedup}. We also notice that profile richness is negative only for a statistically intangible number of routines: this means that \trms-based profiles are (almost) always at least as informative as those obtained using the \rms.

Due to Inequality~\ref{eq:trms-larger}, \trms\ values are always larger that \rms\ values for the same routine activations. Figure~\ref{fig:input-volume} characterizes the increase of the input size values due to induced first-accesses on a representative set of benchmarks. The interpretation of these graphs is similar to Figure~\ref{fig:richness}: a point $(x,y)$ on each benchmark-specific curve means that $x\%$ of routines have input volume $\ge y$. E.g., in benchmark {\tt fluidanimate} roughly $3\%$ of the routines take almost all their input from external devices or from other threads. The trend of curves in Figure~\ref{fig:input-volume} decreases steeply from 100 to 0, reaching its minimum at  $x\simeq 8\%$ for most benchmarks: this means that $8\%$ of the routines are responsible of thread intercommunication and streamed I/O, and the input size of these routines cannot be appropriately predicted by the \rms\ metric alone.

\begin{figure}[t]
\centering{
\hspace*{-2.5mm}
\includegraphics[width=1.05\columnwidth]{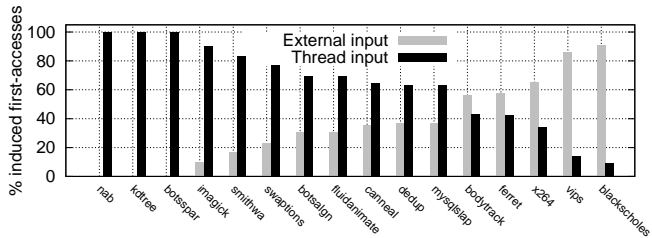} 
}
\vspace{-4mm} 
\caption{External vs. thread-induced input.}
\label{fig:external-vs-thread-hist}
\end{figure}

%--------------------------------------------------------
\paragraph{Analysis of induced first-accesses.} In the previous experiments we observed that a non-negligible number of routines communicate with other threads or with the kernel via system calls. A natural question is how many induced first-accesses are due to external input or are thread-induced. 
Figure~\ref{fig:external-vs-thread-hist} answers this question, plotting the percentage of thread-induced and external input on a representative set of benchmarks: percentages are computed  over the  total number of induced first-accesses, and therefore sum up to $100\%$. Benchmarks are sorted by decreasing percentage of thread-induced input (and thus by increasing external input). An interesting observation is that the SPEC OMP2012 benchmarks get naturally clustered in the leftmost part of the histogram (from {\tt nab} to {\tt botsalgn}), and all of them have thread-induced input  larger than $69\%$. We notice that external input is predominant in {\tt vips}, which seems in contrast with Figure~\ref{fig:thread-vs-external}. This contradiction, however, is only apparent and has a clear explanation. Figure~\ref{fig:thread-vs-external} plots external input on a routine-per-routine basis, while the global percentage in Figure~\ref{fig:external-vs-thread-hist} is routine-independent: the external input of a specific routine also includes the external input of all its descendants in the call tree, which is instead neglected in the global benchmark measure (where each induced first-access is counted only once in the percentage computation). Similar considerations apply to {\tt mysqlslap}.

\begin{figure}[t]
\centering
\begin{tabular}{cc}
\hspace{-3mm}
\includegraphics[width=0.445\columnwidth]{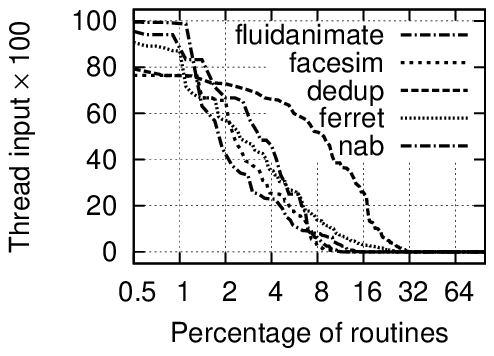} &
\includegraphics[width=0.445\columnwidth]{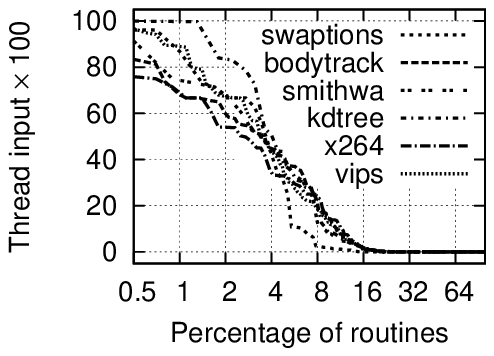}\hspace{-2mm} \\
\hspace{3mm}\small{(a)} & \hspace{6mm}\small{(b)} 
\end{tabular}
\caption{Thread-induced input on a routine basis.}
\label{fig:thread-induced}
\end{figure}

\begin{figure}[t]
\centering
\begin{tabular}{cc}
\hspace{-3mm}
\includegraphics[width=0.445\columnwidth]{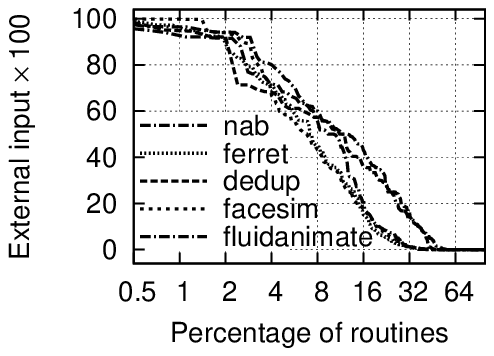} &
\includegraphics[width=0.445\columnwidth]{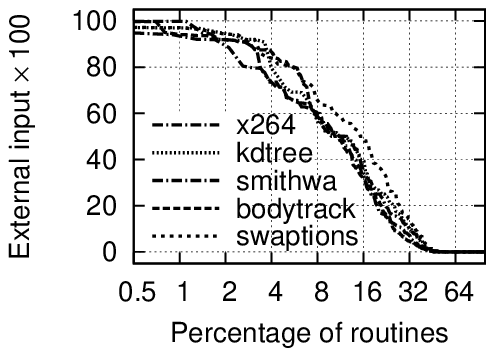}\hspace{-2mm} \\
\hspace{3mm}\small{(a)} & \hspace{6mm}\small{(b)} 
\end{tabular}
\caption{External input on a routine basis.}
\label{fig:external}
\end{figure}

For the sake of completeness, Figure~\ref{fig:thread-induced} and Figure~\ref{fig:external} provide a quantitative evaluation of thread-induced and external input on a routine-per-routine basis: a point $(x,y)$ on each benchmark-specific curve means that $x\%$ of routines have external / thread-induced input $\geq y\%$. For instance, in benchmark {\tt dedup},  $16\%$ of the routines are such that at least $20\%$ of their induced first-accesses are due to thread intercommunication. These charts are in the spirit of Figure~\ref{fig:thread-vs-external}, but exploit a more compact representation.

%We notice that in most benchmarks the percentage of external input appears to be higher than thread input, suggesting that these benchmarks may be heavily multi-threaded, but not parallel in terms of shared memory accesses. We believe that this aspect deserves further investigation, in the spirit of the analysis conducted in~\cite{KMJV12}.

% !TEX root = paper.tex
%--------------------------------------------------------------------------------
\section{Related Work}
\label{se:related}

There is a vast literature on performance profiling, both at the inter- and intra-procedural level: see, e.g.,~\cite{GKM82, ABL97, HG93,  ZSCC06, AR01, HC01, BM05b, VNC07} and the references therein. All these works aim at associating performance metrics to  distinct  paths traversed in the call graph or in the control flow graph during a program's execution. Input-sensitivity issues are instead explored in~\cite{GAW07, CDF12, MM04, ZH12}. Marin and Mellor-Crummey~\cite{MM04} consider the problem of understanding how an application's performance scales given different problem sizes, using data collected from multiple runs with  determinable input parameters. Goldsmith, Aiken, and Wilkerson~\cite{GAW07} also propose to run a program on  workloads of different sizes,  to measure the performance of its routines, and eventually to fit these observations to a model that predicts how the performance scales. The workload size of the program's routines, however, is not computed automatically. Algorithmic profiling by Zaparanuks and Hauswirth~\cite{ZH12}, besides identifying boundaries between different algorithms in a program, infers their computational cost, which is related to the input size. The notion of input size is defined at a high level of abstraction, using different definitions for different data structures (e.g., the size of an array or the number of nodes in a tree). Instead, the input-sensitive profiling methodology described in~\cite{CDF12}, which  provides the basis for our approach, automatically infers the input size by tracing low-level memory accesses performed by different routines. None of these approaches addresses concurrency issues, being thus limited to sequential computations.

The problem of empirically studying the asymptotic behavior of a program has been the target of extensive research in experimental algorithmics~\cite{McGeoch07, McGSFCP02, DFI03}, where individual portions of algorithmic code are extracted from applications and separately analyzed on ad-hoc test harnesses. This approach has some drawbacks as a performance evaluation method in actual software development, most prominently the fact that, by studying performance-critical routines out of their context,  possible performance effects due to  the interaction with the overall application might be missed.

A variety of parallelism-related profilers have been proposed to help programmers parallelize complex codes by uncovering the dependencies between different regions of the program: examples include {\tt pp}~\cite{L93}, Alchemist~\cite{ZNS09}, and Kremlin~\cite{GJLT11}. Other profilers for multicore machines, such as~\cite{MV10, PZM10, LLQ12}, focus on NUMA-related performance issues and exploit detailed temporal information about memory accesses in order to build temporal flows of interactions between threads and objects. This is similar to our problem of relating memory accesses with thread intercommunication, although the final goal is orthogonal to ours, since these works aim at understanding the speed improvements that can result from parallelizing different portions of code, from executing a program on a parallel platform, or from diagnosing and reducing distant memory accesses.

% !TEX root = paper.tex
%--------------------------------------------------------------------------------
\section{Conclusions}
\label{se:conclusion}

In this paper we have extended the input-sensitive profiling methodology to concurrent computations. Input-sensitive profiling requires to measure automatically the size of the input given to a generic code fragment: in a multithreaded scenario, this raises a variety of interesting issues mainly due to thread intercommunication via shared memory. At this aim, we have proposed a novel metric, called \trms, that gives an estimate of the size of the input of each routine activation by taking into account first-accesses, possibly induced by other threads or by kernel system calls. We have shown that our approach is both methodologically sound and practical. Namely, our Valgrind-based implementation achieves performances comparable to other prominent heavyweight analysis tools. As a future direction, it would be interesting to adapt our methodology to a fully scalable and concurrent dynamic instrumentation framework, in order to exploit parallelism to leverage the slowdown of our profiler.

Our methodology raises many interesting open issues regarding input characterization and thread intercommunication in concurrent applications. Measures derived from \trms\ might allow it to evaluate concurrency-related aspects and to discover how multithreaded applications scale their work and how they communicate via shared memory. E.g., in a recent experimental study~\cite{KMJV12}, it has been observed that even widespread multithreaded benchmarks do not interact much or interact only in limited ways, and that communication does not change predictably as a function of the number of cores. This study exploits a characterization of read/write memory accesses, and we believe that the \trms\ might shed new light towards this direction.

%\bigskip
%
%\noindent {\bf CHECKLIST}:
%
%\begin{itemize}
%\item maiuscole vs. minuscole nei titoli dei paragrafi
%\end{itemize}

%----------------------------------------------------------------
\bibliographystyle{abbrvnat}
\softraggedright
\balance
\bibliography{paper}

\end{document}